# Data Preprocessing for Evaluation of Recommendation Models in E-Commerce

**Namrata Chaudhary** [1,*] and **Drimik Roy Chowdhury** [2,*]

[1] Boxx.ai | AI for E-commerce, Data Science dept., Bengaluru 560095, India
[2] University of Michigan, Department of Mathematics, Ann Arbor, USA 48109 & Boxx.ai | AI for E-commerce, Data Science dept., Bengaluru 560095, India
\* Correspondence: namrata@boxx.ai (N.C.); drimikr@umich.edu (D.R.C.);
Tel.: +91-9620878024 (N.C.); +1-(734)-904-9384 (D.R.C.)

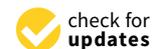

**Abstract:** E-commerce businesses employ recommender models to assist in identifying a personalized set of products for each visitor. To accurately assess the recommendations' influence on customer clicks and buys, three target areas—customer behavior, data collection, user-interface—will be explored for possible sources of erroneous data. Varied customer behavior misrepresents the recommendations' true influence on a customer due to the presence of B2B interactions and outlier customers. Non-parametric statistical procedures for outlier removal are delineated and other strategies are investigated to account for the effect of a large percentage of new customers or high bounce rates. Subsequently, in data collection we identify probable misleading interactions in the raw data, propose a robust method of tracking unique visitors, and accurately attributing the buy influence for combo products. Lastly, user-interface issues discuss the possible problems caused due to the recommendation widget's positioning on the e-commerce website and the stringent conditions that should be imposed when utilizing data from the product listing page. This collective methodology results in an exact and valid estimation of the customer's interactions influenced by the recommendation model in the context of standard industry metrics, such as Click-through rates, Buy-through rates, and Conversion revenue.

## 1. Introduction

The advent of the e-commerce markets facilitates the process of shopping without the need for physical interactions with products. However, an appealing aspect of physical retail stores is that customers who are undecided on the products they desire to purchase have the ability to browse and receive recommendations from shelf displays and salespeople. The e-commerce industry utilizes recommendation models to satisfy this objective [1].

A personalized recommendation model aims to identify products that are of most relevance to a customer based on his or her past interactions. This enhances a user's intention to browse more products and makes them more likely to buy these products, effectively increasing e-commerce revenue [2,3]. Thus, the evaluation of recommendation algorithms for a range of properties is essential in order to select the best algorithm from a set of candidates [4]. Performance is either measured by offline evaluations, by conducting user studies, or by using online evaluations when a recommendation model is live on an online platform [4]. Online evaluations are unique in that they allow direct measurement of overall system goals, such as long-term profit or user retention [4].

Datasets used to evaluate algorithms have a major impact on results and an evaluation of the strength of a recommendation system is only appropriate if the underlying data is exempt from inaccurate or corrupt records [5,6]. Data inconsistencies may arise either from user bias [7], malicious online bots [8,9], or probable data warehousing problems. The preprocessing performed in this paper is important due to the widespread relevance of ensuring data quality and cleanliness because 80% of mining and data analysis is generally attributed to this effort [10]. In our experimental study of performing an online evaluation of a collaborative-based recommendation model on a sample of 10 different types of e-commerce websites that our organization provides services to, including merchandising clothes, jewelry, grocery, and several other niche products, we have identified 12 major contributors of erroneous data when analyzing the online clickstream e-commerce data. Based on the broad sources these errors descended from, we have categorized them as being a result of *customer behavior*, *data warehousing/technical*, or *user-interface issues*.

The customer behavior section focuses on actions performed by visitors that might hinder a true estimation of an evaluation metric, and therefore what the metric represents might not be a consequence of the recommendation model's influence on the visitor [7]. For example, a high bounce rate [11,12] indicates that there exists a segment of customers that do not view or interact with the recommendations provided on a website; including the data from such users would underestimate the recommendation model's strength when computed in terms of Click-through rate or Buy-through rate. Other issues comprise data collected from the presence of B2B buyers [13], outliers [14], new customers on an e-commerce website, and the possibility of recording duplicate records [15].

Data collection issues stem from inaccuracies in the data records that must be handled to avoid an incorrect performance evaluation of the recommendation model. One such case is the difficulty faced in identifying a unique visitor across multiple devices used by that visitor [16]. This is tackled with a robust mapping algorithm, utilizing both the cookie IDs (from multiple devices) and user ID (from signup) of a customer. Other cases include identifying difficulties in uniform data management, online bot removal [8,9], errors in capturing data, and the presence of product combos or bundles [17]. The impact of each of these cases, along with the techniques to resolve them, are detailed and tested across e-commerce websites.

The design and the ease of accessing products on a company's website has a major impact on the online browsing experience and purchases that occur on the website [18]. The user-interface section examines complications arising from improper placement of the recommendation widget on the e-commerce website and specific handling of data collected from the product listing page to avoid overestimating the strength of the recommendation model.

The applications of all the data cleaning procedures mentioned in the paper are tested using the industry standard metrics: Click-through rate (CTR) and Buy-through rate (BTR). The techniques used to handle the erroneous data have been tried and tested using real clickstream data from various types of e-commerce websites and will be outlined as examples in this paper.

Additionally, we introduce a method in Section 5 that validates whether or not the CTR and BTR calculations are actually a result of the recommendations provided on the e-commerce website. The aim of this section is to ensure that customer interactions in terms of clicks or buys on the recommended products are not false positives and a mere result of random click behavior.

## 2. Related Work

*2.1. General Sources of Errors and Data Cleaning Strategies*

In his paper, Hellerstein discusses the sources of erroneous collection of data which stem from "initial data acquisition to archival storage" [14]. The most pertinent sources of errors in the discussion of data preprocessing with regards to user interactions and clickstream activity are:

- Data entry errors: data entries recorded by humans from speech or written context, where most often there are fields of data stored that might not always have a value present; hence an improper

- default value may be assigned to this field without much consideration to whether or not the default value is a possible outcome of real data.
- Distillation errors: data that should be preprocessed before storing them onto a database to reduce complexity and noise of raw data, which if not considered, may impede final analysis. For example, incremental revenue or the conversion metrics calculated may be invalid if the prices utilized for each product varies in the currency they are stored against. It can then be beneficial to assert this when recording data for e-commerce businesses who have a global dominance, to standardize the currency used against one system (such as the USD) for all customer interactions. This notion serves as the background to Section 4.2.4 Uniform Data Management.
- Data Integration errors: when the data collected stems from various servers or sources where the merging of records into a single database may cause inconsistencies. For example, it is crucial that the timestamps of each user interaction on the database are stored against a standard time zone, such as using Coordinated Universal Time (UTC), and not against the time zone of the IP registering the interaction. This is discussed in Section 4.2.4.

Pertinent data cleaning categories involved stem from that of in quantitative data, numerical values that capture some unit of measure and quantities of interest, and identifiers, which corroborate to the need of uniquely identifying objects in the data. Section 4.1.1 on outlier detection and Section 4.2.2 on unique visitor identification serve as examples of the quantitative data and identifier class of data cleaning techniques, respectively [14].

According to Srivastava et al. [10], data preprocessing is defined to "convert the usage, content, and structure information into the data abstractions necessary for pattern discovery" [10]. The content cleaning converts images and texts into understandable mediums to determine usage patterns across websites. The relevance of this aforementioned strategy could be to transform the product identifiers on the e-commerce website into a format comprehensible by the recommendation algorithm and for data analysis, thereby keeping the format consistent throughout several applications [10].

*2.2. Data Records Duplication*

Kohavi et al. [15] discuss a variety of data cleansing challenges that arise from software bugs, implementation, data collection, and transfer process. A significant challenge explored is that of duplicated customer records. Erroneous data warehousing may cause duplicate entries of customer interactions to appear in the collected data. To handle this issue, an efficient way for de-duping [15] is needed, which requires the identification of unique customers. This may pose a challenge because customers may have multiple accounts or one account may be shared amongst multiple customers. Refer to Section 4.2.2 for an effective mapping algorithm for each customer to their unique identity.

Deshmukh and Wangikar [19] discuss an accurate and successful algorithm to detect duplicate records through the idea of "smart tokens", which removes the need to parse extremely long records several times to identify this duplication. The performance has been measured to have very high recall and precision. This technique can be used in the context of simply correcting for duplicated customer interaction (e.g., click, buy) entries in the recorded database [19].

*2.3. Web Bot Detection*

Xu et al. [20] present an effective bot detection approach, which has been tried and tested, to identify a large collection of IPs used by malicious bots; this is conducted by assigning IPs an abnormal score and investigating the reasons behind why some IPs have a score greater than some predefined threshold. The findings have identified a set of characteristics of these bot IP addresses (BIPs) that can be utilized in the data preprocessing stage for such bot identification. A few notable results that are worth mentioning is that 88% of BIPs are only active for a mean value of 1.7 days and often clear cookies up to 100 times a day to avoid their detection. Moreover, the BIPs are stated to not be active during hours 20 to 23, which is when most human users are said to be; most BIPs operate in the working hours of the normal human user. Furthermore, BIPs have been identified to have very aggressive consecutive searches, where the time in between clicks is approximately less than $1/5$ of the average user. Regarding the particular deployment of bots, 99.9% of BIPs were loaded on PC devices, whereas ~20% on wireless or mobile (where the BIP can be on both) [20]. Several other findings and statistical verification of results are explored in [20]. Similarly, Suchacka [21] analyzes the differences of bot and human sessions of traffic on e-commerce websites based on the log files, which provides more insight on the differences in the number of requests made, session duration, mean time per page, etc.

In addition, Tyagi et al. [22] examine a bot detection scheme utilizing web usage mining, which is the "automatic discovery of user access patterns from web servers". The algorithm identifies the robots and web-spiders' navigation patterns by studying the user agent (browsers identify the entity on every request in a field called User-agent) and remote hostname properties of the weblogs. Furthermore, this article explores manners to clean server log entries to reduce the noise and volume of extraneous data in the database [22]. Kohavi also investigates the notion of user agents by identifying those that have never logged on or purchased products, performed only 1 click, or more than 100 click sessions, or with no referrer on all their requests [23].

Lastly, Suchacka et al. [24] outlines a Bayesian approach for detecting web bots that typically hide their identifies from weblogs on servers, which is tested and validated on data from an active e-commerce website, with accuracy greater than 90%.

## 3. Data Description

*3.1. Data Collected*

- A list of products shown to a user as recommendations are recorded along with DateTime and user information and is referred to as a hit.
- A click, add-to-cart, and buy action performed on products on the e-commerce website is recorded along with DateTime and user information.
- Each different web-element on the page of an e-commerce website that displays recommendation products referred to as a widget.
- To identify a unique visitor on the site, the cookie ID of the browser session is stored in the data along with the account username (i.e., user ID) in case the user is logged into his or her account.
- Cust ID is a mapping identifier created (Section 4.2.2) for the purposes of associating activity data under a unique ID using cookie and user ID.

*3.2. Definition of Terms:*

- **Customer** and **visitor** are used interchangeably to refer to an individual on an e-commerce website who performs activities on the website and is provided with recommendations.
- The webpages where widgets are located and personalized products are re-ordered on an e-commerce website are identified by one of the following categories:

    ○   **Home page**: This is the first page visible when visiting an e-commerce site
    ○   **Product List Page (PLP)**: This is the page that re-orders a list of personalized products within a unique department of products (e.g., Home and Kitchen)
    ○   **Product Display Page (PDP)**: This page displays a specific product that was clicked on where similar products based on a defined relation to the viewed product are displayed (e.g., widgets like "You May Also Like", "Similar products")
    ○   **Cart page**: The page on an e-commerce site that displays products that are currently in a customer's purchase cart and might showcase products attributed to the class "People who bought this also bought"

*3.3. Definitions of the Basic Evaluation Metrics*

- **Click-through rate (CTR)** is a ratio that shows how often people who see your products end up clicking it. It can be computed by dividing the number of times a recommended product was clicked by the number of times recommendations are seen. This metric is ideally reflective of the influence personalized recommendations have had on a customer's journey on the website.
- **Add-to-cart-through rate (ATC-TR)** and **Buy-through rate (BTR)** are defined similarly as CTR but using add to carts and buys, respectively.
- **Conversion revenue** is defined to be the total revenue earned through the purchase of recommended products.

A constraint to adhere to when computing the above metrics is that one needs to ensure that recommended products were actually noticed by customers before the users went on to perform an action on the items. A way to enforce this constraint is to introduce a maximum time difference of a valid recommendation-induced-click from the time the recommendations have been shown to when a customer performs an interaction on one of the products recommended.

Based on statistics stated by Kohavi [23], an average visitor spends 5 min and a purchasing visitor spends 30 min on an e-commerce website. Therefore, when computing CTR, ATC-TR, and BTR, we ensure that the corresponding activity occurred within 5 min, 30 min, and 24 h of the recommended product being shown to a visitor, respectively. The reasoning behind the 24 h constraint for BTR is simply to allow the customer sufficient time to consider a purchase decision, which has been explored and tested as a suitable time frame from our recommendation engines implemented.

## 4. Data Cleaning and Validation Methods

In this paper, we identify various issues that obstruct an appropriate evaluation of a personalized recommendation model for an e-commerce platform. For each of the three defined sources of errors in this paper—customer behavior, technical, and user-interface—we identify and delineate methods to recognize and resolve them.

*4.1. Issues Pertaining to Customer Behavior*

4.1.1. Outlier Customers

Outliers in online activity may exist from a few visitors that randomly clicked or bought too many items on a particular day. We regard such customers as outliers, as their activity is not reflective of the model's recommendations, because such interactions could be resulting from bots or bulk buyers.

A point to be noted is that outlier removal is not necessary for an ideal scenario, where the data is recorded over a long period of time and is sufficient in quantity, such that a few outlying points have no overall impact on the evaluation metrics. However, this is not true in most real-life scenarios, as businesses may either have a limited time to observe various recommendation models on their shopping sites or that they receive low traffic of activity, which could make their clickstream data highly sensitive to outliers.

To identify and eliminate such outliers in data, we try to distinguish them from a population of total click and buy activities per day by a customer, within a time range that captures a significant amount of data. The process eliminates activity data associated with such outlier clickers or buyers on a particular day from the calculation of our metrics—CTR and BTR.

A simple yet robust method that can be applied for outlier removal is through a process called Hampel X84 [14]. This procedure relies on applications of the median and median absolute deviation (MAD) for the detection of outliers because of its ideal breakdown point of 50%. The Hampel X84 categorizes an outlier if the data point is greater than $1.4826x$ MADs away from the median, where $x$ is the number of standard deviations away from the mean where the datum would not be considered an outlier. The reasoning behind the factor of 1.4826 is that for a normal distribution, approximately 1 standard deviation from the mean is about 1.4826 MADs [14].

Therefore, an outlier is a customer whose:

$$Total\ Clicks/Buys\ per\ day\ >\ Median_{clicks/buys}\ +\ 1.486 * MAD_{click/buys} \qquad (1)$$

We choose an example e-commerce website data (Case 1) of views and buys per day, from a week's activity on an e-commerce website. We have a population size of 32,000 records of view data aggregated at a customer, day level, and 820 records of buy data aggregated at a customer, day level, arising from approximately 3000 customers interacting on the site in a day. The result of applying the Hampel X84 technique on this data is shown below in Table 1.

Table 1. Results and observations related to Hampel X84 technique on Outlier Removal for Case 1.

|               | Min | Max | Median | MAD  | Outlier-Upper Limit |
|---------------|-----|-----|--------|------|---------------------|
| Views per day | 1   | 460 | 1      | 3.22 | 10.5                |
| Buys per day  | 1   | 24  | 1      | 1.62 | 5.83                |

However, to eliminate a customer who has clicked more than 10 times a day as an outlier seems too stringent of a condition. Following this heuristic, approximately 6% of total customers involved in Case 1 would have been identified as outliers, which may be a concern in terms of the amount of data lost depending on the degree of activity of an e-commerce business. Upon further inspection, we find that our data does not strictly follow a normal distribution (Appendix A), and therefore the above technique is not suitable. To resolve this, we need a method that does not assume that data follows a particular distribution.

We utilize a non-parametric technique which is based on identifying extreme values from a bootstrap-based outlier detection plot, known as Bootlier plot. Analyzing the modality of this plot indicates the presence of outliers: if the plot is unimodal, one may conclude that the data does not contain outliers. This concept of Bootlier Plot is explained in detail by Singh and Xie [25].

The idea is to sequentially cancel observations from the upper tail (possible outliers) and perform bootstrapping to get mean-trimmed means of multiple random samples each time to generate a Bootlier plot [26]. The Bootlier plot is multimodal if the data contains outliers to begin with, and once we reach a point where the plot has a single peak, a well-founded argument can be made that the outliers in the data have been eliminated according to the findings of Singh and Xie [25]. However, a fact to bear in mind is outliers can be heavily masked by the presence of a large proportion of interactions performed by customers who partake in very minimal views or buys (e.g., in Case 1, approximately 49% of the total customers had just 1 view interaction associated with them). In this case, the Bootlier plot will not be multimodal per se, but will be fairly noisy in its distribution. Upon analysis, the Bootlier procedure is still robust in principle, and one should continue removing interactions from the upper tail if the graph is either multimodal or noisy until the plot appears to be a noise-free bell curve, where this heuristic is dependent on the individual user's restraints on how noise-free the distribution should be (which in turn affects the outlier removal limit).

The three considerations to be made in applying the Bootlier plot method are:

1. The parameter *N* is the size of the randomly selected samples with replacement for a chosen number of bootstrapping iterations. It is set depending on the expected number of outliers in the distribution [25].
2. The variable *k* is a trimming factor used to compute mean-trimmed means for each bootstrap sample, which in turn helps make the presence of rare outliers visible in those samples. As we increase *k*, the Bootlier plot advances to a smoother bell curve, and therefore it must be set to an appropriate value depending upon sample size and type of trimming, as suggested by Singh and Xie [25].

3. The iterating factor *i* is the number of times samples of size *N* are randomly generated with replacement to compute mean trimmed mean post each iteration. For our results, we use *i* = 50,000 as a sufficient number of iterations.

We apply the above method on data from e-commerce websites whose population sizes vary from 3000 to 20,000 data points when all view data is aggregated at a customer, day level, and 500 to 1000 data points when all buy data is aggregated at a customer, day level. We expect, at most, 1% of each population to fall under outliers, and therefore fix the value of the parameter *N* = 1% of total population size. Hence when a large number of bootstrapping iterations are performed, it makes it highly probable that at least a single randomly selected sample of size *N* would contain a large proportion of outliers, whose average would be drastically different from the rest of the randomly selected samples. This would result in a fairly noisy or multimodal Bootlier plot and help us detect the presence of outliers in the population. However, to account for small population sizes when performing outlier removal on buys, we adhere to a lower limit of *N* = 50.

The parameter *k* is fixed across all calculations on data from different e-commerce websites, and given that *N* for views per day is much higher than for buys per day, we fix: *kview* = 7 and *kbuy* = 3 [25].

For Case 1 mentioned above, for the given population sizes, *N* is computed to be 320 (from Case 1 population size of 32,000 for views) through the procedure outlined for the outlier removal process of views. Similarly, *N* = 50 (from Case 1 population size of 820 for buys) for the removal of buys. When the Bootlier plot outlier removal technique is applied in this distribution of total views per day for each customer, it results in a unimodal histogram after elimination of data points more than 32 from the population. Similarly, in the case of the distribution of total buys per day, the Bootlier plot approaches unimodality post elimination of data points greater than 14 in the population of buys/day for each customer. This technique classifies 0.8% of unique customers as outliers.

Figures 1 and 2 are examples of an e-commerce website having moderate traffic. Applying the Bootlier process on data from other online businesses exhibiting different levels of traffic, the corresponding outlier limits were seemingly appropriate given the respective context (Appendix B).

- Case 2: A high traffic e-commerce website with ~6000 customers per day resulted in unimodal Bootlier plots after removal of greater than 77 views and 7 buys per day, which classified 0.3% of all unique customers as outliers (Figures A2 and A3).
- Case 3: A low traffic website with ~2000 customers per day (maximum 13 buys) resulted in unimodal Bootlier plots after removal of greater than 19 views and required no removal of buys, which classified 0.1% of all unique customers as outliers (Figures A4 and A5).

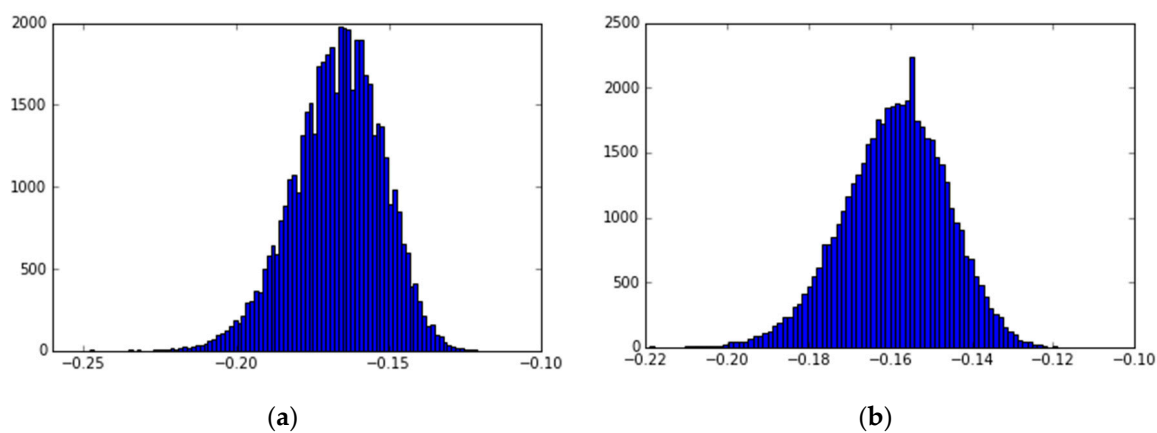

**Figure 1.** Bootlier plot applied for total views/day at customer level for Case 1. (**a**) An initial noisy/multimodal plot before removal of any observations which shows the presence of outliers. (**b**) Final unimodal plot after data points greater than 32 are removed from the population of view data, which is outlier free. (The original data is in the Supplementary Materials).

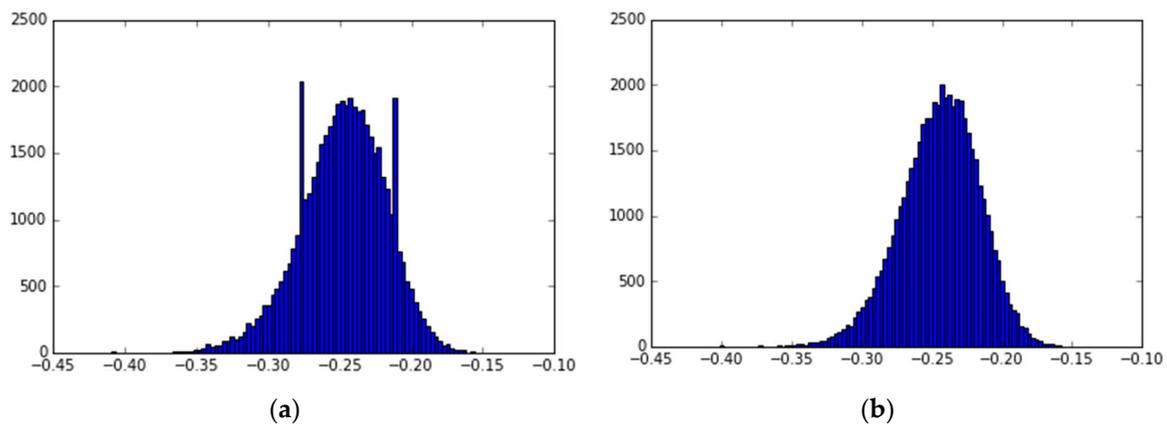

(**a**)　　　　　　　　　　　　　　　　　　　(**b**)

**Figure 2.** Bootlier plot applied for total buys/day at customer level for Case 1. (**a**) An initial noisy/multimodal plot before removal of any observations which shows the presence of outliers. (**b**) Final unimodal plot after data points greater than 14 are removed from the population of buy data, which is outlier free. (The original data is in the Supplementary Materials).

Additional details and characteristics of the three cases of e-commerce websites explored are given in Table A1 of Appendix B.

To summarize, though the Hampel X84 is based on parameters that are robust against the influence of outliers, this procedure has an underlying dependency on a statistical distribution (e.g., Gaussian), which may or may not be valid (Appendix A), and this method has been noted to remove a significant portion of raw data. The non-parametric Bootlier plot method serves as a technique to determine an acceptable demarcation limit for outlier identification regardless of the degree of traffic of an e-commerce business, based on the testing presented above and in Appendix B; however, there may arise the need for the user to manually delineate a threshold of data removal, beyond which the Bootlier plot appears noise-free and unimodal.

4.1.2. B2B Customers

Some e-commerce sites could be a source for retailers to use purchased products for resale or development of other goods. These individuals are referred to as B2B customers, and it has been observed that their characteristics differ from the common visitors in the following aspects: (i) they purchase twice as much as B2C customers [13] and several sites supporting B2B e-commerce provide bulk buying options and bulk pricing [27]; (ii) they purchase rationally compared to possible impulsive buys from B2C customers [28]. In such cases, it becomes very important to identify such buyers specifically, as they are not the target audience for our recommendation model and greatly inflate the purchase data. Furthermore, B2B customers often have a relationship specified with e-commerce businesses on the specific products they need, and these individuals mostly purchase from that standard set of products [29].

Based on the above observations, a way to identify them could be by checking whether a customer engages in more buys than some predefined threshold, customized for each e-commerce website. We define this threshold by computing the median purchases per day for each customer and identify potential B2B customers by segregating the ones that buy $m$ times more products per day than the median value. The value chosen for $m$ can vary based on the type of business, and can be as small as 5 in case of an e-commerce business selling fashion products, or larger than 10 in case of grocery and household items. Note that this suggested value of $m$ is just a heuristic based upon the hypothesis that a reseller would buy 5 to 10 times more items in a single day compared to a regular customer, and can be modified as per one's business case.

In an example we dealt with, where we were aware of the presence of B2B buyers on an e-commerce website selling fashion products, an average purchasing customer bought ~2 items per day. Taking *m* = 5, potential B2B buyers were delineated as the ones that bought more than 10 items per day. In a time period of data ranging over 15 days, 0.03% such customers were identified and eliminated, who contributed to 14% of all purchases.

Given that B2B customers do not buy on impulse and might include multiple individuals [28] taking purchase decisions, once B2B customers have been identified, an alternative to personalization is to segment them under a separate category who are recommended popular products based on some time frame and belonging to the product department the B2B customer intends to purchase from.

4.1.3. Duplicate Online Activity

Duplicate purchases or add to carts of the same item by a customer within a single session can be wrongly used to overestimate the BTR and ATC-TR metric, respectively. This is because technically, our model has led a customer to discover only one unique item. We have usually encountered such duplicates due to the following reasons:

1. Erroneous data gathering process could result in multiple records of the same action being stored with a small difference in timestamp [15]. In this case, it is essential to drop duplicate actions to maintain data integrity.
2. Products that have been bought or added to cart multiple times could be continually recommended because of the large quantity of interactions performed on them (e.g., click, buy) [30]. Dropping duplicates proves to be important here as well, because the recommendation model has led the customer to find only one unique product in his or her exploration. Thus, disregarding such instances is necessary for both model evaluation and training.

From our internal observations of multiple e-commerce websites, those shopping sites selling grocery or household items are often the businesses that have products purchased in bulk by customers.

4.1.4. Visitors Bounced off the Home or Product Listing Page

In the case of e-commerce sites with low traffic, a proportion of their customers might fall in the category that are visiting the website for the first time and leave the site directly from the home page or PLP without performing any interactions [11,12]. The reason for this could either be uninterested visitors or an unappealing website design [31]. The proportion of data associated with these customers is not relevant to the analysis of the recommendation engine because such users have no historical interactions that could have provided them with personalized recommendations. For each of these occurrences, a hit is recorded, but such customers do not intend to shop, and therefore perform no activity. The proportion of such customers arriving every day is called the bounce rate of a website, and the hits for these customers must be removed to keep from deflating CTR and BTR metrics.

For one such e-commerce site internally studied, we analyzed that the bounce rate was 43%, and these bounced customers who only received one hit each and had no other activity data, contributed to 17% of the total hits. Upon the removal of these hits from the observation, the CTR metric increased by 20%, from 6.7% to 8.1%.

Sculley et al. [32] use bounce rates of advertisements as a measure of poor advertiser-return on investment and conclude that a high bounce rate may mean that users have a bad experience post clicking on the particular advert. Therefore, they use this metric to formulate a method that predicts the bounce rate by analyzing the features of the advertisement. This allows them to validate the quality of future advertisements and predict their effectiveness [32].

In a few e-commerce businesses studied internally, we observed that the bounce rate was almost close to 70%. Here, it becomes very important to remove individuals contributing to this issue and re-run the online experiment for a sufficient amount of days to gather significant data for the computation of results. To remove data contributing to the bounce rate, eliminate all customer data that have only ever received one hit on the home page or PLP and performed no interactions (e.g., a click, buy).

4.1.5. Dealing with Initial Clicks of New Customers on a Website

Evaluating the ratio of new customers arriving every day to the total number of customers can provide valuable information on general customer trends on a shopping site. This is because recommendation models work best for customers that have sufficient past click, ATC, and buy data recorded on the e-commerce website, whereas the initial actions of every new customer that visits a site will possibly not result in them being recommended any personalized products. However, once the new customer performs a notable amount of browsing, the recommendation model will be able to provide more personalized recommendations.

Therefore, when these initial clicks of new users are segmented separately and the CTR for these clicks is compared to that of the CTR computed on the rest of the clicks, one can clearly observe a difference, as demonstrated by Figure 3, wherein the metric is much worse for a customer who has minimal past activity, as opposed to having more activity recorded. Figure 3 is an example of such observations for an e-commerce platform, where the CTR for the first click of a new user is viewed separately from the CTR of users with at least one historic activity. Thus, these initial activities by new users may contribute to lowering the overall CTR calculated for the e-commerce website; we suggest separating them out as they do not reflect true personalization by a recommendation model, and instead considering the CTR of users with at least $x$ ($x$ = 1 in the below case) historic activity for the actual evaluation metric.

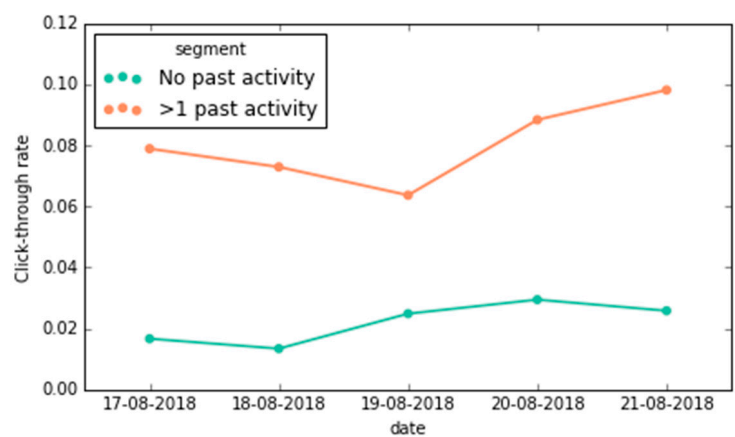

**Figure 3.** This shows that if the overall numbers for all customer data were shown, they would be around 4.5%, but the actual personalization model gives a CTR of around 7–8%.

In the above case, we can note that the CTR results of the personalized model can be more appropriately determined for customers that have at least 1 historic click ($x$ = 1 click). However, this $x$ cut-off limit can vary and can be estimated by incrementally increasing $x$ from 1 click onwards and comparing the CTR below and above that click threshold and observing whether there is a notable difference in the two. The value of $x$ can be fixed once a clear distinction is observed, wherein the CTR of the initial new customer clicks below a cut-off limit $x$ is drastically lower than for the CTR for rest of the data.

Additionally, this case does not apply for BTR, because from our observations, a customer has generally performed enough clicks by browsing products before making a purchase. However, this too might not stand true on the type of website where most customers shop with a utilitarian approach [33], and one may observe new customers purchasing items without a large amount of prior browsing of products.

*4.2. Issues Pertaining to Technical or Data Collection Problems*

4.2.1. Online Bots and Web Crawlers

Web crawlers and shopping bots surf a browser to collect information, compare products between online businesses, and thus, may lead to unwanted interactions occurring on an e-commerce website. Web bots have accounted for an overwhelming 42.2% of all web traffic in a 2018 report [20], with 20% of e-commerce activity being malicious bot activity, which can dramatically change clickstream patterns on a website, in turn skewing any clickstream statistics. Moreover, approximately 70% of Amazon traffic has been attributed to these web crawlers as well [20]. Data resulting from these transactions and hits should not be considered into the calculation of the result metrics, as these interactions occur regardless of what recommendations are shown.

Section 2.3 of this paper discusses a few techniques present in the current literature on how to identify and deal with web bots. We classify bot detection [34] under (i) detection using online signatures and (ii) detection using anomalous activity.

(i) Bots leave specific signatures that can be detected in User Agents and known bot IP addresses. Tan and Kumar [8] list some known User Agents and bot IPs, as well as how to detect ethical bots using HEAD requests (type of request that a web server responds to by sending only the message header) and HTTP requests (a request message sent to a web server on page load) with unassigned referrers.

(ii) Bots compared to human users tend to have very high navigation speeds on an e-commerce website or exhibit some sort of recurring behavior, even with moderate speeds.

   (a) Therefore, one can compute clicks or ATCs per second, and theoretically, a value close to or greater than 1 can indicate an online bot.
   (b) Another technique could be to record and observe recurring patterns in the time difference between two simultaneous clicks or ATCs or effective session durations [24]. Unique users exhibiting such behavior can also be categorized as a bot.

4.2.2. Track a Unique Visitor

The whole basis of the recommendation system present on an e-commerce website lies in its ability to provide products personalized to each customer. If the underlying data is unable to associate interactions on a website correctly with the corresponding user, then an argument in favor of the personalized recommendation system is certainly flawed.

The issue that resides in this identification process is that of retaining a specific cookie ID and a corresponding user ID throughout all of a customer's interactions on the e-commerce website, which is often not plausible due to the following reasons:

I. *Incognito browsing, clearing of cookies, or use of multiple devices.* In either case, a new cookie ID is registered for all following interactions of a customer, thereby disassociating a link to the customer and his past interactions against the older cookie ID. In this case, it would be possible to identify a unique customer through his or her user ID, given the customer has logged in once prior or created an account on the website. However, this cannot be employed on the interactions on a private browsing session when a customer never logged in; hence, this data would, by default, get mapped under a new visitor. Additionally, if a customer were to complete certain transactions under a cookie ID and then log in (thereby registering the user ID to the cookie ID), it is crucial that those pre-login interactions reflected only under the cookie ID are mapped accordingly to the user ID.

II. *A cookie ID is linked to multiple user IDs.* If there are any interactions or hits recorded for a cookie ID with multiple user IDs present across data, then it is possible that multiple individuals are using the same browser on a device. In such cases, it is impossible to identify the true customer behind an action when the customer is not logged in and we advise that such cases should be eliminated.

To handle these aforementioned cases, the following algorithm has been provided to clean the original data.

**Algorithm:**
Let **Data** be the dataset containing all hits and interactions
Let **Map** be a unique mapping of all cookie IDs and user IDs created from **Data**
Pseudocode

1. Eliminate entries from Data where both cookie ID and user ID are NULL/Missing
2. Eliminate entries from Data where both user ID is NULL/Missing and the cookie ID in the record has multiple unique user IDs as indicated by Map
3. For each row in Data:
    Create a field called Cust ID
        If user ID $\neq$ NULL:
            b. Cust ID = user ID
        Else:
            If cookie ID found in Map:
                c. Cust ID = user ID in Map $_{cookie\ ID}$
            Else:
                d. Cust ID = cookie ID

Initially in step 2, using the dataset Map, we detect all cases where a cookie ID is mapped to multiple user IDs, and remove all data associated with those cookie IDs where a user ID is missing, because these are cases where it is not possible to identify the associated user ID amongst the individuals logged on previously onto the e-commerce website. In step 3, if the algorithm detects that the user ID field is null for a particular row, this deals with all the records from when a visitor is not logged into his account, causing the user ID field to be empty, and we treat this by backfilling the Cust ID field with a user ID that has been associated with the cookie ID found in the dataset Map. If cookie ID is, in fact, not present in Map, this means that the cookie ID was never associated with a user ID, because the particular user never logged in on the website. In such cases we can continue identifying this user by his or her unique cookie ID; therefore, the Cust ID is assigned as the cookie ID of that visitor.

Campbell [16] describes an approach that Google Analytics uses to keep track of a unique customer across the different devices used by the customer. Usually, unique visitors are tracked by an identifier Client ID, which is a unique ID that Google Analytics generates from a web browser and device combination. The source of error is that this Client ID is lost once the user switches to another device. As a solution, they introduce User ID, an ID stored based on a customer's login into Google and suggest User ID to be used as a replacement for Client ID to better track a user across devices [16]. The approach we discuss above is built upon a similar concept, where we use an algorithm to create a mapping called Cust ID from the combination of a browser's cookie and login user ID. As a result, our method additionally helps with backtracking and mapping all data with a unique customer, resolving issues with incognito browsing, clearing of cookies, or use of multiple devices, and identifying problematic cases where a cookie ID is linked to multiple user IDs.

We worked with an e-commerce site where the above algorithm played an important role of mapping transactions to a unique visitor across user IDs and cookie IDs, which resulted in an improvement of 11% in CTR, increasing it from 1.54% to 1.71%. This was possible because there were cases of undetected clicks, where a cookie ID was missing but user ID was present that was later identified and used in the CTR calculation. An important note is that cases where a cookie id was associated with multiple user IDs—the set of interactions that were considered invalid for analysis, as per our method—accounted for 0.2% of all cookie IDs in the raw data. Hence, the general evaluation conducted on customer behavior is not affected significantly when disregarding these entries in data.

4.2.3. Error in Capturing Online Activity Data

An ideal journey of a customer shopping for a product on an e-commerce website is:

$$\text{Click} \rightarrow \text{Add to Cart} \rightarrow \text{Buy}$$

The purpose of checking underlying data for an ideal customer journey on each product is that such an inspection can give us insight into the accuracy of the data warehousing process. In case we were to find a group of customers whose click was missing even when an add to cart was present for a product, or similarly, click or add to cart was missing when a purchase is present, this could be indicative of an integration error, wherein the process is failing to capture some type of actions for these customers. Additional analysis on these customers could indicate them using a particular type of device or browser where the data integration is failing, and this insight could be very helpful in correcting that underlying integration and warehousing mistake. A presence of large-scale unaccounted data in an ideal journey could be the result of data integration issues that arise from dynamic websites and structural changes of an e-commerce website [35].

In terms of the evaluation using the available data, if the errors present in an ideal customer journey are not a significant constituent of the overall data, removing all interaction data for such customers on the particular products would possibly lead to a more reliable calculation of our evaluation metrics. Otherwise, analyzing the root reason behind these errors is clearly the priority.

On a side point, if it has been validated that data is being collected correctly and yet various buys or add to carts are occurring without a click, this could be a result of the fact that few e-commerce businesses bypass a stage in the ideal journey—for example, through the employment of "Quick buy", a product is added to cart and ready to be purchased without recording a click. In these cases, there isn't any error present and the process of evaluation can be carried out without any eliminations. Hence, the check for such a condition also must be performed in the UI/UX of the shopping site before searching for a reason for missing interactions or modifying the evaluation method.

4.2.4. Uniform Data Management

This idea is fairly straightforward and self-explanatory in its purpose, in the sense that all data stored should be uniform and consistent (e.g., a unit of measure). For example, an issue that may arise for global e-commerce businesses is that the prices recorded for each product sale might not be standardized and stored against one currency. These types of issues are categorized as common *distillation errors* [14], as mentioned in Section 2.1. This is particularly alarming in the case of calculating conversion revenue, as the summation of monetary values would be of products considered in various currencies. Another issue that could transpire is if the data recorded are not all kept to a standard time zone. Possibly, individuals in different time zones on an e-commerce website could have their respective hits and transactions recorded on separate servers that have system clocks not registering these activities to one uniform time zone upon saving the data to a unified location [14]. This could result in erroneous calculations of the evaluation metrics when observed over a time frame.

4.2.5. Dealing with Activity on Combo/Bundle Products

Retailers can choose to market and sell certain products in combinations as a bundle because these products express their true valuation more efficiently for an entire bundle than for a single item [17]. In our research and testing on a wide range of e-commerce sites, we encountered a business that sells products in combos—for example, "a family of four", "mom and child", "brother and sister"—for groups of clothes. These are not marketing bundles, but a particular characteristic of the business itself. Due to this modification in the type of retailing a business performs, the method to deal with the backend data of such e-commerce sites also differs, and one such case is presented in this section.

When products are clicked on and added to cart on a site that sells items in a combo, the data stored against these actions record the product bundle using its combo identifier, but when a purchase occurs, each item belonging to that combo is recorded by its individual SKU in the data. This type of data mismanagement is more prevalent where a business retails almost all its products in the form of a combo. In case of such data, an evaluation would result in an accurate computation of the CTR and ATC-TR, but will inflate the computation of BTR as multiple buys are recorded for each single product combo sold (which, in this case, is equivalent to one collective product).

To resolve this, one could use the mapping of individual combo item SKUs to their respective combo IDs, and therefore compute BTR based on the unique combos that were purchased by each customer.

However, in some cases, this mapping might not be available, and to this, we propose an alternate metric to the BTR that measures the likeliness of a customer to buy products (combos in this case) on the e-commerce site. It can be computed as:

$$BTR_{BUYER} = (\text{Number of unique buyers who received recommendations})/(\text{Number of hits}).$$

*4.3. Issues Associated with the User-Interface of the Website*

4.3.1. Visibility of Recommendation Widgets

An accurate personalized recommendation model is only effective when correctly implemented on the user-interface of the target webpages. Forrester Research estimates that "poor web design will result in a loss of 50% of potential sales due to users being unable to find what they want, and a loss of 40% of potential repeat visits due to initial negative experience" [18].

In our observations from multiple e-commerce sites, we noticed cases where the reason for low user interaction was not due to irrelevant products that were recommended to users, but a result of recommendations not directly visible on opening a page. Seroussi states in his article [36] "You can have the most accurate recommendations in the world, but no one would know about it (or care) if they are not served in a timely manner through a friendly interface".

Such cases generally occur when, for example, the widget on a PDP (product display page) is not placed close to or immediately under the currently open product and requires the user to perform some amount of scrolling to view the widget. Another scenario is when the website loading speed is too slow to display the recommended items within a short time before a user moves on to a newer page.

As a result, a large proportion of visitors do not even glance at the products personalized for them, and the effect of personalization is limited. In such cases, we may observe very low CTR and BTR rates on specific pages of a website irrespective of displaying relevant products as recommendations.

4.3.2. Design of the PLP

The PLP is designed to display products based on a categorical filter (e.g., shirt or trousers, or color of shirts). A PLP that uses a recommendation engine would display products re-ordered in the decreasing order of "most likely to be bought or clicked on" (personalization) by a visitor, and this reordering is an output of the recommendations provided by the model, which usually occurs for the remainder of the pages that can be browsed on this PLP search. This notion of the decreasing order of relevance of products on PLP sites is further employed in practice and suggested as optimization of personalization [37].

Therefore, a stringent condition is added for performance metric calculations where only the top *N* products (mostly top 4–8 products on the top 2 rows) visible on the initial display (visible at default zoom on a browser page without requiring a scroll) on page 1 of the PLP are advised to be the only PLP hits taken into account for the validity tests. This is because the accuracy of a recommendation model should be reflective from the initial products a customer views on a page by default without having to scroll or switch to a newer page.

Additionally, we have empirically observed that not applying such a condition inflates the CTR of a PLP, making it drastically higher than the CTRs of other web pages, for the same recommendation model, customer, and e-commerce products, which raises doubt on the evaluation result of that PLP page.

## 5. Method to Validate Effectiveness of a Recommendation Model

This section aims to provide a method that can be used to validate if recommended products actually have an impact on what a customer views and purchases on an e-commerce site. Through this process, we try to confirm if a measured CTR or BTR for a recommendation model is actually reflective of the percent of products a customer finds useful, and not some random number generated as a result of that customer's random clicks or predetermined buys.

The method is based on the concept of A/A Testing in an online experimental setup [38,39]. This type of experiment divides users into two segments, and then both are assigned identical models (hence the name). The aim is to test if there is truly a lift that can be attributed to using the model and check if we are not just observing a false positive resulting from natural variation. Similarly, in our case with a similar setup, we test if, for example, an $x$% CTR is actually the result of users satisfied with $x$ clicks out of every 100 recommendations shown to them and not a result of random clicks or predetermined buys, which is a natural activity on a shopping site.

Therefore, we split all customers randomly into two equal sections that are both shown product recommendations using the same model and uniquely recognized by a flag. Therefore, the data recorded for customers under each section will be marked with the respective flag of the section they fall in, and this method of recording data makes it easier to compute CTR and BTR for the two sections.

If a comparison between the CTR or BTR of the two sections shows very little difference and high correlation over days, then we may conclude that the result metrics are not random occurrences and that the customers are impacted by personalized recommendations. If not, the evaluation metrics do not reflect the recommendations' influence on a customer for that particular website, and therefore, are futile in the analysis of the recommendation engine. The results of such a test are indicated below in Figure 4 for the metric CTR.

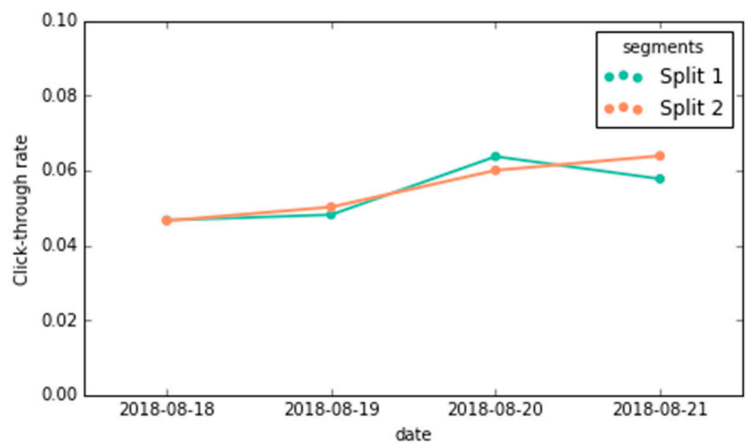

**Figure 4.** Shows an example of such a test on an e-commerce website data, with 50/50% random split of customers under Segment 1 and 2 provided with the same model; it can be observed from the values that the difference between Click-Through Rates is negligible, as well as similar over days.

## 6. Conclusions and Evaluation

In a world of burgeoning e-commerce and online transactions, the application of recommendation platforms is increasingly valuable to facilitate better trade. Continuous improvement of these recommendation models can be ensured by learning its flaws through tests and evaluations on live e-commerce websites. However, such evaluations and online clickstream data, in general, are prone to numerous errors, and therefore, we discuss 12 such errors that stem from three sources: customer behavior-based, due to data warehousing and technical issues, and user-interface issues.

The first section explores the issue of customer behavior-based errors with regards to problems such as outlying customers who perform an exceptionally high number of actions on a website that cause inflated metric calculations because of their metrics' sensitivity to these extreme values. As a solution, we propose the non-parametric technique of Bootlier plots to identify a limit of views or buys per day for each customer, beyond which a data point can be categorized as an outlier. This technique is validated with the clickstream data from 3 cases of different e-commerce websites.

Furthermore, the presence of a few B2B buyers on regular e-commerce websites has been noted to exaggerate the evaluation metrics as these customers are typically not susceptible to the personalized recommendations for reasons outlined in Section 4.1.2. We suggest the removal of such buyers from the underlying data before performing an evaluation of a personalized recommendation model, and the paper provides a technique to do so along with a real-life e-commerce business in consideration. Presence of duplicate data records is shown along with the reasons in Sections 2.2 and 4.1.3; a regular de-duping of data is suggested to avoid duplicates. Furthermore, we point out how visitors bouncing off an e-commerce website (Section 4.1.4) aren't a negative effect of the recommendations suggested on that website, and how the removal of the hits by these customers from data can better evaluate the model, as high bounce rates may severely deflate the CTR and BTR metrics from their true population value. Lastly, we explore how a model fails to personalize some initial recommendations to a new user on a website in Section 4.1.5. As a workaround, we propose that the first few clicks of such new customers should be removed from the evaluation, and an approach to do so is proposed.

The second section explores technical issues ranging from the presence of bots, challenges of tracking a unique visitor on websites, and loss of data due to integration problems. Additionally, the warehousing challenges of collecting and maintaining uniform data and recording purchase data for product bundles are discussed (Section 4.2.4). Techniques to identify bot actions by the means of either online signatures or anomalous activity are suggested (Section 4.2.1). For unique visitor identification (Section 4.2.2), an algorithm is devised for the purpose of creating a unique identifier, called Cust ID, which is a result of using cookie IDs and user IDs of online visitors. This helps map all actions underlying data correctly under this identifier.

Finally, under technical issues, we explore how a deviation from the ideal journey of a customer shopping for a product (Section 4.2.3) can indicate problems with the underlying integration system that collects data, and that the data for such customers with an alternative to the Click → Add to Cart → Buy sequence must be ignored for a true performance evaluation. Under the data warehousing section, we propose using a clear mapping between the ID of a product combo and the individual product IDs within each combo (Section 4.2.5). This can help eliminate the issue where earlier purchases were counted multiple times for each product ID within a combo. If such a mapping is absent, then a replacement metric is suggested for the standard BTR.

The third section in our paper discusses some problems related to the UI/UX presentation of recommendations on an e-commerce website. The first problem explored is the low visibility of recommendations on a PDP or home page due to poor placement or slow loading of widgets (Section 4.3.1). We suggest these issues be fixed by the development team of the website. The second issue is in the calculation of CTR and BTR on a PLP page, which should be limited to considering the hits only from the first few selected number of products from the recommendations shown on the first page (Section 4.3.2). This facilitates a more efficient evaluation of the recommendation model by being able to verify if it is able to influence a customer's clicks within the few initial and most relevant products displayed.

In the last section (Section 5), we also suggest that performing A/A tests on the e-commerce website ensures that the validation received from customers through clicks or buys in a CTR or BTR evaluation, respectively, is actually a result of them being satisfied with the recommendation model and not a metric accidentally generated through random clicks and predetermined buys.

Recommendation engines enhance the customer's knowledge of relevant products on the e-commerce website and potentially attribute a greater inflow of revenue for the business in question. Hence, both customers and businesses are viably benefitted from the inclusion of the recommendation platforms. To keep improving the recommendation models, the preprocessing techniques discussed allow for a more veracious evaluation of the true state of the recommendation influence on customers and businesses and improve the merits that such recommendation engines can deliver.

## 7. Limitations

The non-parametric Bootlier plot explored in Section 4.1.1 with regards to outlier removal has been presented in such a way that it might, in some cases, require the analyst to manually inspect and estimate whether or not the plot appears to have a noise-free and unimodal distribution. Moreover, the added variability of whether or not the plot appears to have these aforementioned characteristics can be ambiguous in terms of execution, and there is certainly an element of personal judgment in deciding the outlier limit. In Section 4.2.2, because of the restrictions placed in determining unique user identities, certain customers who have never logged on and keep changing their devices or keep clearing cookies cannot be identified across data perfectly. Next, the strategy to identify B2B users might not be sufficient to segregate out each and every B2B customer, as the method has been tried on a limited number of e-commerce businesses, and given it is only a heuristic approach, one cannot ensure the identification of every B2B customer.

## 8. Future Work

A few of the cases, such as the section of B2B customers where we utilize a simple heuristic approach in order to identify such individuals, have not been explored in a depth necessary to formulate a robust method that can tackle the presence of each error. Moreover, the establishment of these 12 issues is an attempt to a provide extensive coverage and a comprehensive guide for proper data preprocessing, but to achieve this, an in-depth analysis has been compromised for some of the problems. Therefore, in the future, some of the more pressing issues presented in this paper can be explored profoundly.

Section 4.1.1 on outlier removal is an issue which depends on the particular activity (view, ATC or buy) distribution of an e-commerce website inspected. In the future, more cases should be explored for this outlier detection strategy of the Bootlier plot on the different departments of e-commerce businesses (grocery, jewelry, etc.) and on differing degrees of customer activity (in terms of interactions performed on average). Hence, there could be a holistic view of the certain preprocessing techniques that are more applicable for certain types of e-commerce businesses, as compared to others.

Even though these data preprocessing techniques have been outlined in terms of their responsibilities and goals, it is valuable to discern what exact information presented from the cleaned evaluation metrics, such as CTR and BTR, indicate on the appropriateness of the recommendation algorithm and customer behavior.

**Supplementary Materials:** The following are available online at http://www.mdpi.com/2306-5729/4/1/23/s1.

**Author Contributions:** Conceptualization, N.C. and D.R.C.; methodology, N.C. and D.R.C.; validation, N.C. and D.R.C.; formal analysis, N.C. and D.R.C.; investigation, N.C. and D.R.C.; resources, N.C. and D.R.C.; data curation, N.C. and D.R.C.; writing—original draft preparation, N.C. and D.R.C.; writing—review and editing, N.C. and D.R.C.; visualization, N.C. and D.R.C.; project administration, N.C. and D.R.C.

**Acknowledgments:** A special thanks to the organization, Boxx.ai | AI for e-commerce for the material, administrative and technical support. Specifically, we would like to appreciate Aman Sohane (Data Science lead at Boxx.ai) for his input on the conceptualization. We would also like to thank Rajat Mhetre (Data Scientist at Boxx.ai) and Ajay Kashyap (Co-founder at Boxx.ai) for their help with computations and paper structure, respectively.

**Conflicts of Interest:** The authors declare no conflict of interest. The funders had no role in the design of the study; in the collection, analyses, or interpretation of data; in the writing of the manuscript, or in the decision to publish the results.

**Appendix A**

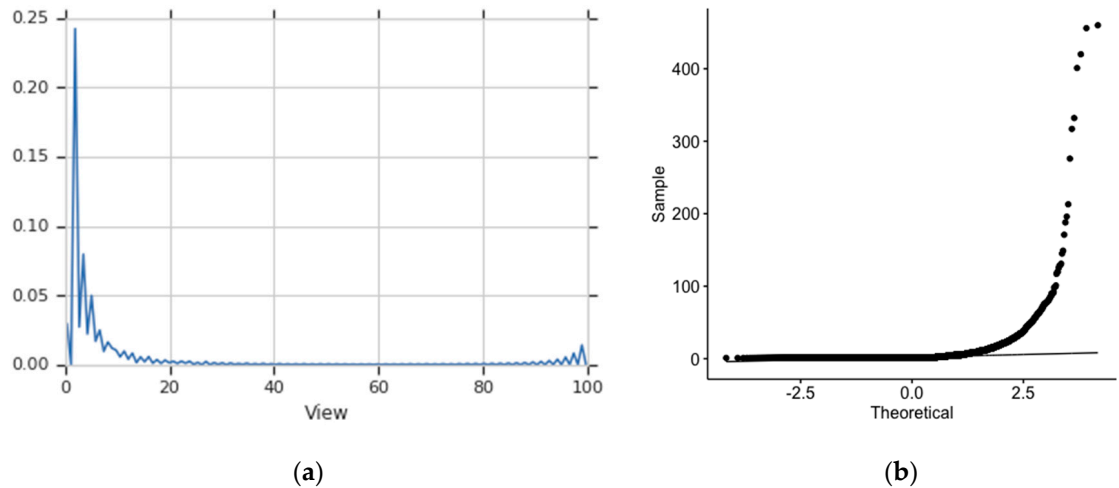

(**a**)  (**b**)

**Figure A1.** Represents the check for normality on views per day data from Case 1. (**a**) A density plot on the data ** contains a large tail and therefore is non-normal. (**b**) A Q-Q Plot, which also results into a non-normal distribution. (** The data used for density plot has been filtered for less than 100 view per day to allow for a better visual of the peak which otherwise was not visible due to the very large tail in data). (The original data is in the Supplementary Materials).

**Appendix B**

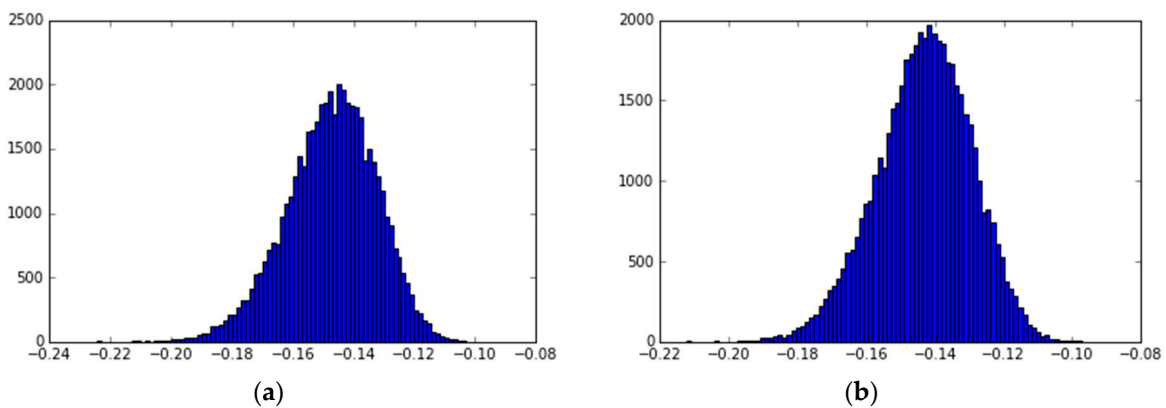

(**a**)  (**b**)

**Figure A2.** Bootlier plot applied for total views/day for each customer for Case 2. (**a**) An initial noisy/multimodal plot before removal of any observations which shows the presence of outliers. (**b**) Final unimodal plot data points greater than 77 are removed from the views' population, which is outlier free. (The original data is in the Supplementary Materials).

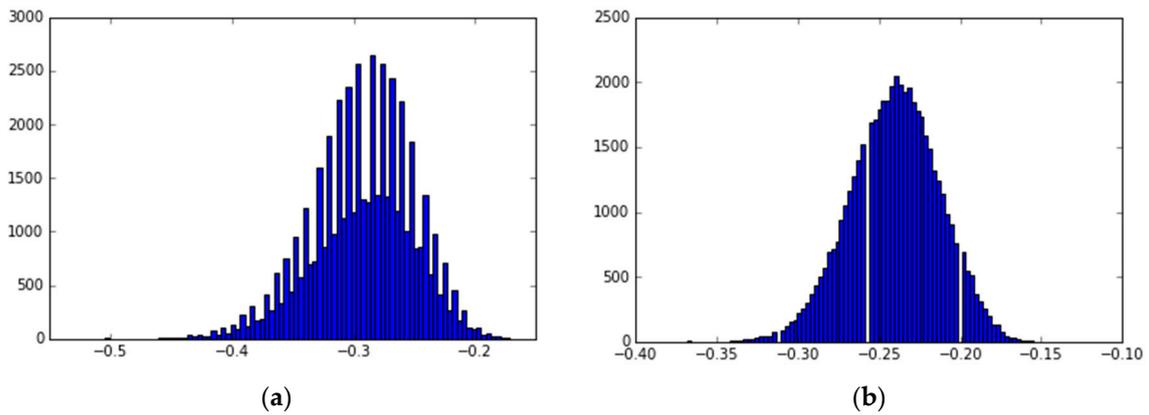

**Figure A3.** Bootlier plot applied for total buys/day for each customer for Case 2. (**a**) An initial noisy/multimodal plot before removal of any observations which shows the presence of outliers. (**b**) Final unimodal plot data points greater than 7 are removed from the buys' population, which is outlier free. (The original data is in the Supplementary Materials).

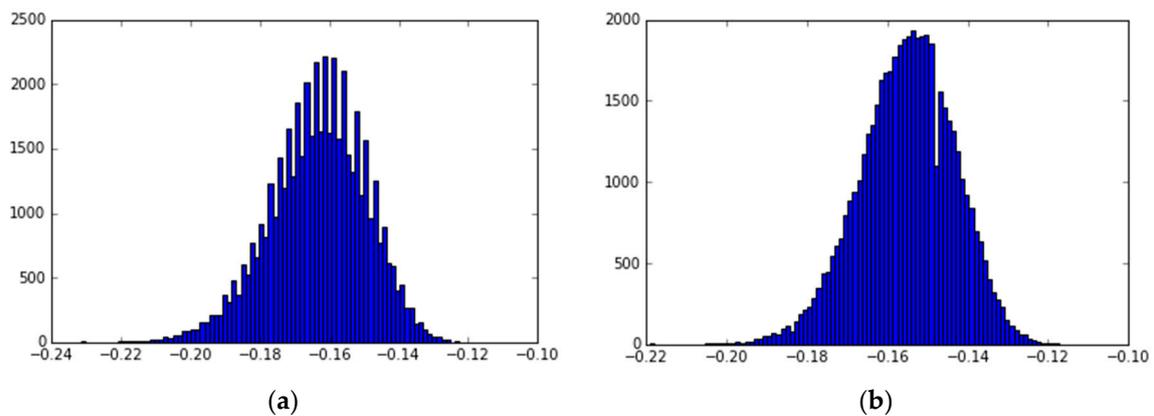

**Figure A4.** Bootlier plot applied for total views/day for each customer for Case 2. (**a**) An initial noisy/multimodal plot before removal of any observations which shows the presence of outliers. (**b**) Final unimodal plot data points greater than 19 are removed from the views' population, which is outlier free. (The original data is in the Supplementary Materials).

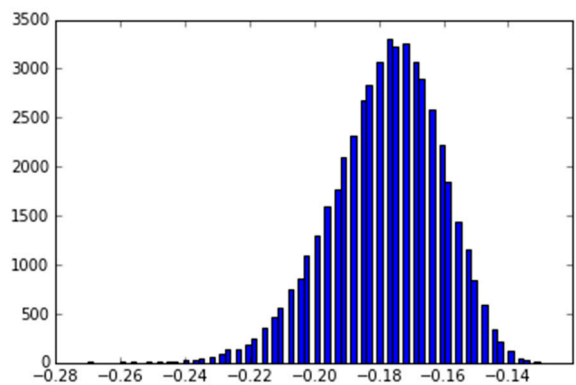

**Figure A5.** Bootlier plot applied for total buys/day for each customer for Case 3. The initial unimodal plot shows that all buys are outlier free to begin with. (The original data is in the Supplementary Materials).

**Table A1.** E-commerce website characteristics of Case 1, 2, and 3 in Section 4.1.1. (The original data is in the Supplementary Materials).

| Cases | Products Type | Total Customers | Time Period | Max Buys/Day by a Customer | Max Views/Day by a Customer |
| --- | --- | --- | --- | --- | --- |
| Case 1 | Fashion clothing and accessories | 29,357 | 10 days | 24 | 460 |
| Case 2 | Traditional Indian wear | 31,767 | 5 days | 20 | 382 |
| Case 3 | Plus-size fashion clothes | 19,111 | 10 days | 13 | 361 |